\documentclass[aps,prl,twocolumn]{revtex4}

\begin{document}


\title{Modified gravity and the stability of de Sitter space}


\author{Valerio Faraoni}
\email[]{vfaraoni@cs-linux.ubishops.ca}
\affiliation{Physics Department, Bishop's University\\
Lennoxville, Qu\`ebec, Canada J1M~1Z7
}


\date{\today}

\begin{abstract}
Within the context of modified gravity and dark 
energy scenarios of the accelerated universe, we study the 
stability of de Sitter space with respect to inhomogeneous 
perturbations using a gauge--independent formalism. In modified 
gravity the stability condition is exactly the same that one 
obtains from a homogeneous perturbation analysis, while the 
stability condition in scalar--tensor gravity is 
more restrictive.
\end{abstract}


\maketitle



The recent discovery that the expansion of the universe is 
accelerated, obtained by studying type Ia supernovae \cite{SN}, 
and the cosmic microwave background experiments showing that the 
universe has nearly critical density \cite{CMB},
call for a theoretical explanation. Two classes of models are 
predominant in the literature: the first class assumes that 
there 
is a form of {\em dark energy} (or {\em quintessence}) 
unclustered at all scales, that accounts for 70\% of the energy 
density $\rho$ of the universe. This dark energy, of unknown 
nature, is necessarily exotic: to generate acceleration in 
Einstein gravity it must have  negative pressure 
$P_{DE}<-\rho_{DE}/3$. The best fit to the observational data 
favours an even more exotic dark energy ({\em phantom energy} or 
{\em superquintessence}) with equation of state parameter $ 
w\equiv P_{DE}/\rho_{DE}<-1$, which is evolving in 
time \cite{w}. Were a value $w<-1$ to be confirmed by 
observations, it 
can not be explained by general relativity with a canonical 
scalar field $\phi$, which is the most common model of dark 
energy, because of the Einstein--Friedmann equation 
$\dot{H}=-\kappa\left( P_{\phi}+\rho_{\phi} 
\right)/6=-\kappa\dot{\phi}^2/2 \leq 0$, which is incompatible 
with $P_{\phi}<-\rho_{\phi} $ (equivalent to {\em 
superacceleration} $\dot{H}>0$). To 
model an equation of state parameter $w<-1$, a phantom field 
\cite{phantom} or 
a field coupled nonminimally to 
the Ricci curvature $R$ (see Refs.~\cite{mybook,NMC} for reviews)  
have been considered. These theories can 
be seen as special cases of scalar--tensor gravity, described by 
the action \cite{ST}
\begin{equation} \label{1}
S=\int d^4 x \sqrt{-g}\left[ \psi(\phi)R-\frac{1}{2} 
\nabla^c\phi\nabla_c\phi-V(\phi) \right] \;.
\end{equation}

The second class of models does not require the presence of 
exotic dark energy but modifies gravity at large scales by 
introducing non--linear (in $R$) corrections to the 
Einstein--Hilbert Lagrangian which become dominant only at late 
times (low curvatures) \cite{MG1}--\cite{MG3}. These theories 
often suffer from problems with the post--Newtonian limit 
\cite{ChibaPLB03,DolgovKawasaki03,PPN} or from instabilities 
\cite{instabilities}, and 
are not 
yet accepted as completely viable theories, but they are 
nevertheless interesting as the cosmic acceleration that we are 
observing may be the first sign of a departure from Einstein's 
gravity. Furthermore, these models are motivated by certain 
compactifications of M--theory \cite{Mtheory}.

In both dark energy models and modified gravity, depending on the 
model adopted, the universe may 
accelerate forever or end its existence at a  finite time in the 
future in a Big Rip or sudden future singularity 
\cite{BigRip,myBigRip,footnote1}. Such singularities have been 
classified 
in Ref.~\cite{NO2}; according to this classification, it is known 
that singularities of type~I can occur in these 
models \cite{myBigRip}, but singularities of other types are not 
excluded.

The fate of the universe depends on 
the presence and size of the attraction basins of attractor 
solutions in the phase space. In many models of both dark energy 
and modified gravity, a de Sitter attractor solution is found 
\cite{deSitterattractors}. 
In this paper we address the issue of the stability of de 
Sitter 
space in modified gravity and scalar--tensor theories. 
It is 
straightforward to  assess stability with respect to homogeneous 
(time--dependent only) perturbations; however, it is more 
significant to establish whether de Sitter space is also stable 
with 
respect to more  general inhomogeneous (space-- and 
time--dependent) perturbations. This is a much more difficult 
task because of the gauge--dependence problems associated with 
this kind of perturbations \cite{Bardeen}. In the following we 
show that for 
modified gravity the stability condition obtained with a 
gauge--independent inhomogeneous perturbation analysis reduces 
to that obtained in the much simpler homogeneous perturbation 
analysis, whereas this is not the case for scalar--tensor 
gravity.

We begin from the generalized gravity action
\begin{equation} \label{2}
S=\int d^4x \, \sqrt{-g} \left[ \frac{f \left( \phi, R 
 \right)}{2}\, -\frac{\omega( \phi)}{2}\, 
\nabla^c\phi\nabla_c\phi 
-V(\phi) \right] \;,
\end{equation}
which contains scalar--tensor gravity as the case $f\left( 
\phi, R \right)=\psi(\phi) R $, and modified gravity $f(R)$ when 
the scalar  field $\phi$ is absent and $f_{RR} \neq 0$. In the 
spatially 
flat Friedmann--Lemaitre--Robertson--Walker (FLRW) metric
\begin{equation} \label{3}
ds^2=-dt^2+a^2(t) \left( dx^2+dy^2+dz^2 \right) \;,
\end{equation}
the field equations are
\begin{eqnarray} 
&& H^2  =  \frac{1}{3F} \left( \frac{\omega}{2} \, \dot{\phi}^2 
+\frac{RF}{2} -\frac{f}{2} +V  -3H\dot{F} 
\right)   \;, \label{4} \\
&& \nonumber \\
&& \dot{H} =   - \, \frac{1}{2F}  \left( \omega \dot{\phi}^2 
+ \ddot{F}  -H\dot{F} \right)  \;, \label{5} \\
&& \nonumber \\
&&  
\ddot{\phi } +3 H \dot{\phi} +\frac{1}{2\omega} \left(
\frac{d\omega}{d\phi} \,  \dot{\phi}^2 - \frac{\partial 
f}{\partial \phi} +2\, \frac{dV}{d\phi}  
\right) =0 \;,\label{6}
\end{eqnarray}
where $F\equiv \partial f/\partial R$, $H\equiv \dot{a}/a$, 
and an overdot denotes differentiation with respect to 
the comoving time $t$. By choosing $\left( H, \phi \right) $ as 
dynamical variables, the equilibrium points of the dynamical 
system (\ref{4})--(\ref{6}) are de Sitter spaces with constant 
scalar field $\left( H_0, \phi_0 \right) $. These solutions 
exist subject to the conditions
\begin{eqnarray} \label{7}
&& 6H_0^2 \, F_0 - f_0+2V_0 =0 \;, \\
&&\nonumber \\
&& {f_0}'- 2{V_0}' =0 \;, \label{8} 
\end{eqnarray}
where $ 
F_0 \equiv F \left( \phi_0, R_0 \right) $, $ f_0\equiv f
\left( \phi_0, R_0 \right) $, $ V_0 \equiv  V  \left( \phi_0 
\right) $, $ {V_0}' \equiv  \left. \frac{dV}{d\phi}  
\right|_{ \phi_0 } $, a prime denotes differentiation with 
respect to $\phi$, and $ R_0 = 12 H_0^2 $.

Inhomogeneous perturbations of de Sitter space are investigated 
by using the covariant and gauge--invariant formalism of 
Bardeen--Ellis--Bruni--Hwang \cite{Bardeen} in the version 
studied by Hwang \cite{Hwang} for generalized gravity. The 
metric  perturbations are defined by 
\begin{eqnarray} 
g_{00} & = & -a^2 \left( 1+2AY \right) \;, \label{9} \\
&& \nonumber \\
g_{0i} & = & -a^2 \, B \, Y_i  \;, \label{10} \\
&& \nonumber \\
g_{ij} & =& a^2 \left[ h_{ij}\left(  1+2H_L  \right) +2H_T \, 
Y_{ij}  \right] \;,\label{11}
\end{eqnarray}
where the scalar harmonics $Y$ are the eigenfunctions of the 
eigenvalue problem $
\bar{\nabla_i}\bar{\nabla^i} \, Y =-k^2 \, Y $. 
Here $h_{ij} $  is the three-dimensional metric of the FLRW 
background and the operator $ \bar{\nabla_i} 
$ is the covariant derivative associated with  $h_{ij}$, while 
$k$ is an eigenvalue. The vector and tensor 
harmonics $Y_i$ and $Y_{ij}$ obey 
\begin{equation} \label{13}
Y_i= -\frac{1}{k} \, \bar{\nabla_i} Y \;, \;\;\;\;\;\;
Y_{ij}= \frac{1}{k^2} \, \bar{\nabla_i}\bar{\nabla_j} Y 
+\frac{1}{3} \, Y \, h_{ij} \;.
\end{equation}
Bardeen's gauge--invariant potentials 
\begin{eqnarray} 
&& \Phi_H = H_L +\frac{H_T}{3} +\frac{ \dot{a} }{k} \left( 
B-\frac{a}{k} \, \dot{H}_T \right) \;, \label{15} \\
&& \nonumber \\
&& \Phi_A = A  +\frac{ \dot{a} }{k} \left( B-\frac{a}{k} \, 
\dot{H}_T \right)
+ \frac{a}{k} \left[ \dot{B} -\frac{1}{k} \left( a \dot{H}_T 
\right)\dot{}  \right] \;,  \label{16}
\end{eqnarray}
and the Ellis--Bruni variable 
\begin{equation} \label{17}
 \Delta \phi = \delta \phi  +\frac{a}{k} \, \dot{\phi}  \left( 
B-\frac{a}{k} \, \dot{H}_T 
\right) \;.
\end{equation}
are used, with equations similar to eq.~(\ref{17}) defining the 
gauge--invariant variables $ \Delta f$, $\Delta F $,  and 
$\Delta R$.  The first order equations obeyed by the 
gauge--invariant perturbations are given in Ref.~\cite{Hwang} 
and they simplify considerably in the de Sitter background 
$\left(H_0,\phi_0 \right)$. To first order, they are: 
\begin{widetext}
\begin{equation}  \label{42}
\Delta \ddot{\phi} + 3H_0  \Delta \dot{\phi} 
+ \left[ \frac{k^2}{a^2}-\,  \frac{1}{2\omega_0} \left( f_0''   - 2 V_0'' \right) \right] 
\Delta \phi  =
 \frac{ f_{\phi R}}{2 \omega_0 } \,  \Delta R  \; ,
\end{equation}

\begin{equation}  \label{43}
\Delta \ddot{F} +3H_0 \, \Delta \dot{F} +\left( \frac{k^2}{a^2} - 4H_0^2 \right) \Delta F 
+\frac{F_0}{3} \, \Delta R =0 \;,
\end{equation}

\begin{equation} \label{44}
\ddot{H}_T +3H_0  \, \dot{H}_T +\frac{k^2}{a^2} \, H_T=0 \;,
\end{equation}

\begin{equation} \label{45}
-\dot{\Phi}_H+H_0 \Phi_A =\frac{1}{2} \left( \frac{\Delta \dot{F}}{F_0} -H_0 \, \frac{ \Delta 
F}{F_0} 
\right) \;,
\end{equation}

\begin{equation}  \label{46}
\Phi_H  = -  \frac{1}{2} \, \frac{ \Delta F}{F_0}   \; ,
\end{equation}

\begin{equation}   \label{47}
\Phi_A + \Phi_H =  - \frac{\Delta F }{F_0} \; ,
\end{equation}

\begin{equation}  \label{48}
\ddot{\Phi}_H + 3H_0 \dot{\Phi}_H  - H_0  \dot{\Phi}_A -3H_0^2 \Phi_A  
 =  - \frac{1}{2}\frac{  \Delta \ddot{F}}{F_0} - H_0 \, \frac{\Delta \dot{F}}{F_0} 
+ \frac{3H_0^2 }{2} \, \frac{\Delta F}{ F_0 }   \; ,
\end{equation}
with 
\begin{equation} \label{49}
\Delta R=6 \left[ \ddot{\Phi}_H + 4H_0 \dot{\Phi}_H + 
\frac{2}{3} \frac{k^2}{a^2} \, \Phi_H 
 -H_0 \dot{\Phi}_A + \left( \frac{k^2}{3a^2} -4H_0^2 \right) 
\Phi_A \right] \;,
\end{equation}
\end{widetext}

Furthermore, vector perturbations do not 
have any effect to first order in the absence of ordinary matter 
\cite{Hwang} and de Sitter space is always stable with respect 
to first order tensor perturbations, as can be seen from 
eq.~(\ref{44}), so we only need 
to worry about scalar perturbations (see Ref.~\cite{deSitter} 
for details). We first consider modified gravity theories 
obtained by 
setting $\phi \equiv 1$ and $f=f(R)$ with $f_{RR}\neq 0 $ in 
the action (\ref{2}). The gauge--invariant perturbations are 
related by  \cite{deSitter}
\begin{eqnarray} 
\Phi_H & = & \Phi_A =-\, \frac{\Delta F}{2F_0}  \label{18} \;,\\
&&\nonumber \\
\Delta R & = & 6 \left[ \ddot{\Phi}_H + 3H_0 \dot{\Phi}_H  + 
\left( \frac{k^2}{a^2}  -4H_0^2 \right) \Phi_H 
 \right] \;, \label{19}
\end{eqnarray}
where $a=a_0 \mbox{e}^{H_0 t}$. By using the fact that 
$ \frac{\Delta F}{F_0}=\frac{f_{RR}}{F_0}\, \Delta R$, one 
obtains $\Delta R=-\frac{2F_0}{f_{RR}}\, \Phi_H $. The 
perturbations $\Phi_H$ and $ \Phi_A$ evolve according to
\begin{equation} \label{20}
\ddot{\Phi}_H+3H_0 \dot{\Phi}_H+\left( 
\frac{k^2}{a^2}-4H_0^2 +\frac{F_0}{3f_{RR}}\right) \Phi_H=0 \;;
\end{equation}
at late times the term $k^2/a^2$ can be safely neglected and 
stability  is achieved if the coefficient of $\Phi_H$ in the 
last term of the left hand side of eq.~(\ref{20}) is 
non--negative, i.e. (using eq.~(\ref{7})), if
\begin{equation}\label{200}
\frac{ F_0^2-2f_0 f_{RR}}{F_0 f_{RR}}\geq 0 \;.
\end{equation}
The spatial dependence of the inhomogeneous perturbations is 
encoded in the eigenvector $k$ of the spherical harmonics; the 
fact that the only term containing $k$ (or the physical wave 
vector $k_{phys}=k/a$) in eq.~(\ref{20}) 
becomes negligible on a de Sitter background implies that 
the spatial dependence effectively disappears from the analysis. 
Eq.~(\ref{200}) coincides with the stability condition that can 
be obtained by a straightforward homogeneous perturbation 
analysis of eqs.~(\ref{4}) and (\ref{5}). Hence, in the 
stability analysis of de Sitter space in modified gravity 
theories one can safely neglect inhomogeneous perturbations and 
limit oneself to the much more approachable homogeneous 
perturbations, thus bypassing the gauge--dependence problems. 
However, this conclusion could not be drawn {\em a 
priori} but it necessarily relies 
on the inhomogeneous perturbation analysis presented. Further, 
this result has been shown to be true only for the stability of 
de Sitter space and not for different attractor solutions that 
may be present in the phase space.

Naively, the physical reason for this considerable formal 
simplification 
could be looked for in the fact that, during the 
quasi--exponential expansion of the universe, inhomogeneities 
(and anisotropies \cite{Waldtheorem}) are redshifted away; this 
is not the whole story though, because the simplification found 
for modified gravity does not occur in scalar--tensor theories. 
In fact, in this case, eqs.~(\ref{44}), (\ref{18}), and 
(\ref{19}), together with 
\begin{equation} 
\frac{\Delta F}{F_0}=\frac{f_{\phi R}}{F_0}\, \Delta R \;,
\end{equation}
yield
\begin{equation} \label{89}
\Delta \ddot{\phi} + 3H_0 \, \Delta \dot{\phi} 
+ \left[  \frac{k^2}{a^2}-\,  \frac{  \left( \frac{f_0''}{2}   -  
V_0'' +\frac{6 f_{\phi R}^2}{F_0} \, 
H_0^2 \right)}{\omega_0  \, \left( 1+\frac{ 3f_{\phi 
R}^2}{2\omega_0 F_0} \right)}   \right] 
\Delta \phi  = 0  
\end{equation}
if $ 1+3f_{\phi R}^2/( 2 \omega_0 F_0) \neq 0$.  The 
stability condition of de Sitter space in scalar--tensor 
gravity then becomes \cite{deSitter}
\begin{equation}\label{100}
\frac{  \left( \frac{f_0''}{2}   -   V_0'' +\frac{6 f_{\phi 
R}^2}{F_0} \, H_0^2 \right)}{\omega_0 \, \left( 1+\frac{ 
3f_{\phi R}^2}{2\omega_0 F_0} \right)} \leq 0 \;.
\end{equation}
For general scalar--tensor theories, this condition is more 
restrictive than the corresponding stability condition obtained 
from a straightforward homogeneous perturbation analysis of 
eqs.~(\ref{4})--(\ref{6}), which is 
\begin{equation}  \label{101}
\frac{ \frac{f_0''}{2}   -   V_0''}{ \omega_0}  \leq 0 \;.
\end{equation}
However, for scalar--tensor theories of the form
\begin{equation}
S=\int d^4x \, \sqrt{-g} \left[ \phi R 
-\frac{\omega(\phi)}{\phi} \nabla^c\phi\nabla_c \phi -V(\phi) 
\right]
\end{equation}
with a single coupling function, eqs.~(\ref{100}) and 
(\ref{101}) coincide \cite{Mike}. The reason for the failure of 
these 
equations to coincide in the general case can be traced to the 
right hand side of eq.~(\ref{42}), in which perturbations 
$\Delta R$ act as a source for the perturbation $\Delta \phi$
(roughly speaking, perturbations $\delta H$ in the Hubble 
parameter, and their derivatives, source scalar field 
perturbations $\delta \phi$): such a term is absent in the 
homogeneous perturbation analysis of the Klein--Gordon equation 
(\ref{6}). 
Analogously, such a term is absent in eq.~(\ref{20}) obeyed by 
the gauge--independent perturbations $\Phi_H $ (or $\Phi_A$) in 
modified gravity. This shows that although modified gravity is 
mathematically equivalent to a scalar--tensor theory 
\cite{TeyssandierTourrenc,ChibaPLB03}, the corresponding 
physics is not completely equivalent.

As an example of application of the stability condition 
(\ref{200}), let us consider the theory described by 
\cite{oneoverR,ChibaPLB03,MG3}
\begin{equation}   
f(R)=R-\, \frac{\mu^4}{R} \;.
\end{equation}
The de Sitter space must satisfy the condition 
$R_0=12H_0^2=\sqrt{3}\, \mu^2$ and the stability condition 
(\ref{200}) can never be satisfied: this de Sitter space is 
always unstable. This situation can be ameliorated by a quadratic 
correction -- in the theory with
\begin{equation} \label{last}
f(R)=R-\, \frac{\mu^4}{R} +aR^2 \;,
\end{equation}
the condition for the existence of de Sitter space is again 
$R_0=\sqrt{3}\, \mu^2$; by applying the stability condition 
(\ref{200}) one obtains that de Sitter space is stable if
\begin{equation}
a> \frac{1}{ 3\sqrt{3} \, \mu^2}
\end{equation}
and unstable otherwise (in particular for negative $a$). 
This result agrees with those of Refs.~\cite{NO1}, which follow 
from an independent analysis of the effective potential in the 
Einstein 
conformal frame version of the scalar--tensor theory equivalent 
to the action~(\ref{2}) with $f(R)$ specified by 
eq.~(\ref{last}). Furthermore, the theory described by 
corrections in both $1/R$ and $R^2$ has much better chances of 
passing the Solar System tests than the theory based on simple 
$1/R $ corrections \cite{NO1}.

A more 
complete discussion of the relation between stability 
of modified gravity and of scalar--tensor theories, and the 
application of eqs.~(\ref{200}) and (\ref{100}) to other specific 
dark  energy and modified gravity  scenarios will be presented 
elsewhere.

\begin{acknowledgments}
This work was supported by the Natural Sciences and Engineering 
Research Council of Canada ({\em NSERC}) and by a grant 
from the Senate Research Committee of Bishop's University.
\end{acknowledgments}


\begin{thebibliography}{99}
\bibitem{footnote1} In certain scenarios, matter dominates again 
at late times and the universe eventually enters a decelerated 
era; however, it is very difficult  to cross the barrier 
$w=-1$ from the region $w<-1$ \cite{crossing}.
\end{thebibliography}

\end{document}